# Distributed Adaptive LMF Algorithm for Sparse Parameter Estimation in Gaussian Mixture Noise

Mojtaba Hajiabadi and Hossein Zamiri-Jafarian
Electrical Engineering Department
Ferdowsi University of Mashhad
Mashhad, Iran
mhajiabadifum@gmail.com

*Abstract*— **A distributed adaptive algorithm for estimation of sparse unknown parameters in the presence of nonGaussian noise is proposed in this paper based on normalized least mean fourth (NLMF) criterion. At the first step, local adaptive NLMF algorithm is modified by zero norm in order to speed up the convergence rate and also to reduce the steady state error power in sparse conditions. Then, the proposed algorithm is extended for distributed scenario in which more improvement in estimation performance is achieved due to cooperation of local adaptive filters. Simulation results show the superiority of the proposed algorithm in comparison with conventional NLMF algorithms.**

## I. Introduction

Adaptive Filters are using in large applications to endow a system with learning and tracking abilities, especially when the signal statistics are unknown and are expected to vary with time. Over the last several years, a wide range of adaptive algorithms has been developed for diverse demands such as channel equalization, spectral estimation, target localization, and interference cancellation. One group of the basic adaptive algorithms is gradient-based algorithms such as the least mean square (LMS) algorithm. The well-known LMS algorithm is perhaps one of the most familiar and widely used algorithms because of its good performance in many circumstances and its simplicity of implementation. The books [1], [2] should provide an excellent sense of the main concepts in this area.

However, the LMS algorithm is a popular method for adaptive parameter estimation, in many scenarios parameters of unknown systems can be assumed to be sparse, containing only a few large coefficients spreaded among many small ones. Using such prior information about the sparsity of unknown parameters can be helpful to improve estimation performance, but standard LMS filters do not exploit such sparsity information. In the past years, several algorithms have been proposed for sparse adaptive filtering using LMS, which was motivated by recent progress in compressive sensing [3]. The basic idea of these techniques is to introduce a penalty into the cost function of the standard LMS to notice sparsity. This achieves better performance than that of the standard LMS for sparse models [4].

Many approaches for signal processing problems have been studied when the additive noise process is modeled with Gaussian distribution. However, for many real-life situations, the additive noise of the system is found to be dominantly nonGaussian. Some examples of nonGaussian environments are the ocean acoustic noise and the urban radio-frequency (RF) noise. Also, in processing of radar and sonar signals we need to deal with nonGaussian noise [5]. When the additive noise process is nonGaussin, LMS algorithm has a poor performance. In [6], it was shown that for some environments with nonGaussian noise, Least Mean Fourth (LMF) algorithm outperforms LMS Algorithm. One of the main drawbacks of the LMF algorithm is, its stability problem. In [7] Eweda proposed NLMF algorithm that overcomes the stability problem.

It is clear that using a distributed estimation approach, by cooperation between local adaptive filters, provides spatial diversity that results in better performance in comparison with local adaptive filter [8],[9]. In this paper, after modifying NLMF adaptive filter for sparse parameter estimation, it is extended for distributed scenario when additive noise of system is Gaussian mixture noise. Computer simulation results show that our proposed adaptive algorithm achieves better performance compared to the conventional adaptive NLMF algorithms.

This paper is organized as follows. After the introduction, sparse NLMF algorithm for estimation of sparse problems in the presence of nonGaussian noise is developed in section II. A novel distributed sparse NLMF algorithm for a finite mixture of Gaussian noise is proposed in section III. Finally, Simulation and comparison results are given in section IV, followed by conclusions in section V.

## II. Adaptive Sparse NLMF Algorithm

Assum that signal $X(n)$ is the input of the system in Fig. 1, with sparse unknown parameters column vector

$$W^o = [w_1, w_2, ..., w_N]^T \quad (1)$$

that $(.)^T$ represents the transpose operator. An observation of output signal is

$$d(n) = X(n)^T W^o + z(n) \quad (2)$$

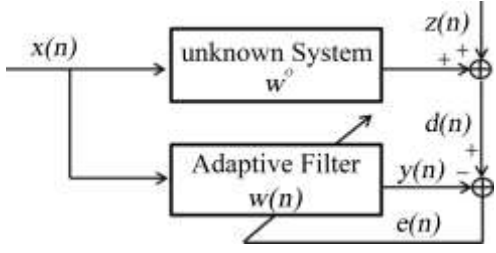

Fig. 1. The block diagram of FIR adaptive filter.

where $X(n) = [x(n), x(n-1), ..., x(n-N+1)]^T$ denotes the vector of input signal and $z(n)$ is the observation noise assumed to be independent of the input signal $x(n)$. The objective of the algorithm is to estimate the sparse unknown vector, $W^o$, using the input signal $X(n)$ and desired output signal $d(n)$. Let $W(n) = [w_1(n), w_2(n), ..., w_N(n)]^T$ be the estimated vector of the adaptive filter at iteration $n$. In the standard LMF, the cost function $J(n)$ is defined as

$$J(n) = \frac{1}{4} e^4(n) \quad (3)$$

where $e(n)$ is the instantaneous error determined as

$$e(n) = d(n) - y(n) \quad (4)$$

in which $y(n)$ is the output of adaptive filter and it is equal to $y(n) = X(n)^T W(n)$. The filter coefficients vector is then updated by

$$W(n+1) = W(n) - \mu_f \frac{\partial J(n)}{\partial W(n)} \\ = W(n) + \mu_f e^3(n) X(n) \quad (5)$$

The stability of (5) is dependent on the statistics of the input and noise signals. In [7] Eweda proposed Normalised version of LMF algorithm, which solves the stability problem of LMF algorithm, as follows:

$$W(n+1) = W(n) + \frac{\mu_f \, e^3(n) X(n)}{\|X(n)\|_2^2 \left( \|X(n)\|_2^2 + e^2(n) \right)} \quad (6)$$

in which, the normalizing term is fourth-order in the regressor and second-order in the estimation error, make the algorithm to be stable against the increase of input power and noise power, respectively. The step-size $\mu_f$, controls the transient and steady-state behavior of the algorithm. It is well-known that, $\|.\|_2$ denotes Euclidean norm operator. Equation (6) can be rewritten as NLMS with variable step size $\mu_f(n)$,

$$W(n+1) = W(n) + \frac{\mu_f e^2(n)}{\|X(n)\|_2^2 + e^2(n)} \cdot \frac{e(n) X(n)}{\|X(n)\|_2^2} \\ = W(n) + \mu_f(n) \cdot \frac{e(n) X(n)}{\|X(n)\|_2^2} \quad (7)$$

To exploit the unknown parameters sparsity, a new cost function, $J_{sparse}(n)$, is defined by combining the instantaneous fourth-order error with the $\ell_0$-norm penalty,

$$J_{sparse}(n) = \frac{1}{4} e^4(n) + \lambda_f \|W(n)\|_0 \quad (8)$$

where, $\lambda_f$, is a regularization parameter, which represents a trade off between estimation error and sparsity of the parameters. Operator $\|.\|_0$ denotes zero-norm operator, which counts the number of nonzero coefficients of vector $W(n)$. Since solving zero-norm algorithm is a hard problem, the zero norm is generally approximated by a continuous function [4]. A popular approximation is:

$$\|W(n)\|_0 \simeq \sum_{i=1}^{N} (1 - e^{-\beta |w_i(n)|}) \quad (9)$$

which leads to the following gradient column vector:

$$\frac{\partial \|W(n)\|_0}{\partial W(n)} = \beta . diag\{e^{-\beta|w_1(n)|}, ..., e^{-\beta|w_N(n)|}\} . sgn(W(n)) \quad (10)$$

in which $diag\{.\}$ represents a diagonal matrix and $sgn(.)$ is a sign function. Using the gradient descent updating, the sparse NLMF filter update is given as

$$W(n+1) = W(n) \\ + \frac{\mu_f e^3(n) X(n)}{\|X(n)\|_2^2 \left( \|X(n)\|_2^2 + e^2(n) \right)} + Sparse\ Penalty \quad (11)$$

By exerting sparse penalty to the standard NLMF cost function, the solution will be sparse and the gradient descent recursion will accelerate the convergence rate of near-zero coefficients in the sparse system. Equation (11) can be rewritten as follows

$$W(n+1) = W(n) + \frac{\mu_f e^3(n) X(n)}{\|X(n)\|_2^2 \left( \|X(n)\|_2^2 + e^2(n) \right)} \\ - \frac{\mu_f}{\|X(n)\|_2^2 \left( \|X(n)\|_2^2 + e^2(n) \right)} \lambda_f \frac{\partial \|W(n)\|_0}{\partial W(n)} \quad (12)$$

In the next section, the proposed distributed algorithm is presented based on (12) as an update equation for each local adaptive filter.

III. ADAPTIVE DISTRIBUTED SPARSE NLMF ALGORITHM

It is well-known that using a distributed estimation approach provides spatial diversity that results in better performance in comparison with local adaptive filter [8]. In this section, we propose a novel distributed sparse adaptive algorithm to further improve the filtering performance in the presence of a Gaussian mixture noise. Obviously, the effectiveness of any distributed scenario will depend on cooperation that are allowed among the nodes [9]. Here, we consider a network with $K$ adaptive filters that the goal of each one is to estimate the unknown sparse vector, $W^o$. The $k$th adaptive filter has access to local data $\{d_k(n), X_k(n)\}$ and the sparse unknown vector, $W^o$, relates to the local data as

$$d_k(n) = X_k(n)^T W^o + z_k(n) \quad (13)$$

where $z_k(n)$ is Gaussian mixture noise that is white in both space and time. In diffusion strategy [8], as shown in Fig. 2, each node $k$, cooperates with its neighborhood nodes. $N_k$, is defined as a set of nodes linked to node $k$, including $k$ itself. In this way, node $k$ combines its *local estimate*, $W_k(n)$ with its neighbors' estimates, $\{W_\ell(n), \ell \in N_k\}$, as follows,

$$\phi_k(n) = \sum_{\ell \in N_k} c_{\ell k} W_\ell(n) \quad (14)$$

The coefficient $c_{\ell k}$ denotes the weight of cooperation, which is chosen as,

$$c_{\ell k} = \begin{cases} \dfrac{1}{|N_k|} & , \ell \in N_k \\ 0 & , \ell \notin N_k \end{cases} \quad (15)$$

where $|N_k|$ is defined as the size of set $N_k$. If we define local error signal as:

$$e_k(n) = d_k(n) - \phi_k(n)^T X_k(n) \quad (16)$$

Based on (12), the distributed algorithm recursions can be given as:

$$\begin{aligned} W_k(n+1) &= \phi_k(n) \\ &+ \frac{\mu_f}{\|X_k(n)\|_2^2 \left(\|X_k(n)\|_2^2 + e_k^2(n)\right)} e_k^3(n) X_k(n) \\ &- \frac{\mu_f}{\|X_k(n)\|_2^2 \left(\|X_k(n)\|_2^2 + e_k^2(n)\right)} \\ &\times \lambda_f \beta.\mathrm{diag}\{e^{-\beta|\phi_{k_1}(n)|},...,e^{-\beta|\phi_{k_N}(n)|}\}.\mathrm{sign}(\phi_k(n)) \end{aligned} \quad (17)$$

Local Mean Square Deviation (MSD) is defined as a performance criterion,

$$V_k(n) = E\|W^o - W_k(n)\|^2 \quad (18)$$

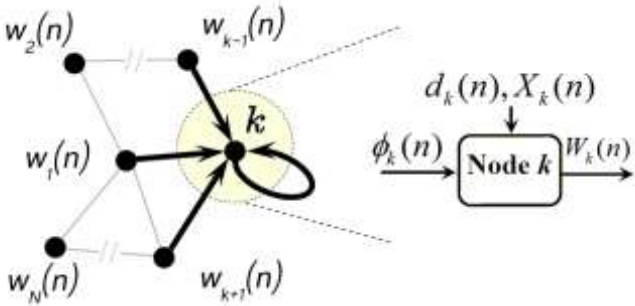

Fig. 2. Distributed network consist of local adaptive filters.

For comparing the performance of the proposed distributed adaptive algorithm (17) with local adaptive filters described in the previous section, we average the local MSD, $V_k(n)$, over the nodes for each time iteration $n$, and so total MSD is defined as,

$$V(n) = \frac{1}{N}\sum_{k=1}^{N} V_k(n) \quad (19)$$

## IV. COMPUTER SIMULATIONS

In this section, we present our simulations of adaptive algorithms described in the previous sections. For all of the simulations, the input signal $x(n)$ is assumed to be white Gaussian process with unit variance. Sparse parameters vector is chosen as :

$$W^o = [0, 0.9, 0.03, 0.7, 0.01, 0, 0.09, 0, 0, 0.01, 0, 0, 0.01, 0, 0.015, 0]^T, \quad (20)$$

that was shown in Fig. 3. The sparsity ratio of $W^o$ is equal to 2/16 which means vector, $W^o$ contain only 2 large coefficients. The observation noise, assumed to be independent in both space and time, is modeled as a Gaussian mixture defined by [10],

$$f_Z(z) = \sum_{i=1}^{3} a_i N(z \mid \mu_i, \sigma_i^2) \quad (21)$$

that was shown in Fig. 3, where the weights are all positive and sum to one.

$$\sum_{i=1}^{3} a_i = 1, \quad a_i \geq 0 \quad (22)$$

The parameters of Gaussian mixture noise are given in the Table 1. For simulation of (17) we consider a network consist of $K = 10$ nodes and $|N_k| = 4$, as shown in Fig. 4.

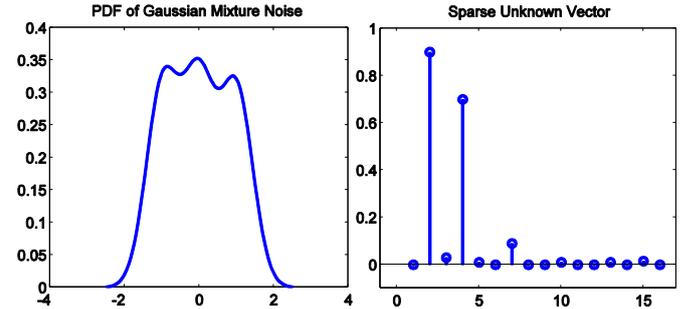

Fig. 3. Probability density function of Gaussian mixture noise (left), and Sparse parameters vector (20) with sparsity ratio 2/16 (Right).

TABLE I. GAUSSIAN MIXTURE MODEL

| Gaussian Number | Gaussian Mixture Components | |
|---|---|---|
| | *Mean* $\mu_i$ | *Variance* $\sigma_i^2$ |
| $i = 1$ | -1 | 0.01 |
| $i = 2$ | 0 | 0.01 |
| $i = 3$ | 1 | 0.01 |

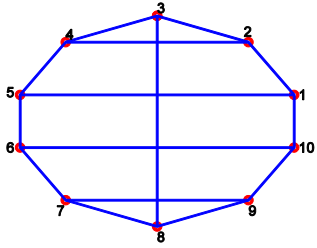

Fig. 4. Distributed network with *K*=10 nodes and |*N*$_k$| = 4.

The performance of the proposed adaptive sparse NLMF algorithms evaluated by computer simulations are shown in Fig. 5, and Fig. 6, for local and distributed scenario, respectively.

In Fig. 5, the local adaptive NLMF algorithm has been simulated based on (12) for Gaussian mixture noise. As seen, our proposed NLMF has superior performance in comparison with NLMS algorithm both in convergence rate and steady state estimation error power.

Fig. 6, shows the performance of the proposed distributed adaptive NLMF algorithm simulated based on (17). As seen, the distributed NLMF algorithm outperforms other local adaptive algorithms regarding convergence rate and MSD criteria.

## V. CONCLUSIONS

A distributed adaptive Normalized Least Mean Fourth (NLMF) algorithm for sparse estimation in Gaussian mixture noise has been proposed in this paper. At first step, local adaptive NLMF algorithm has been modified for sparse estimation by zero norm criterion in order to speed up the convergence rate and also to reduce the steady state estimation error power. Then, the proposed algorithm has been extended to distributed scenario in which more improvement in performance has been achieved due to spatial diversity gain. It should be note that although the proposed distributed algorithm has a better performance in both transient and steady states, the computational complexity of distributed algorithms is higher than local ones.

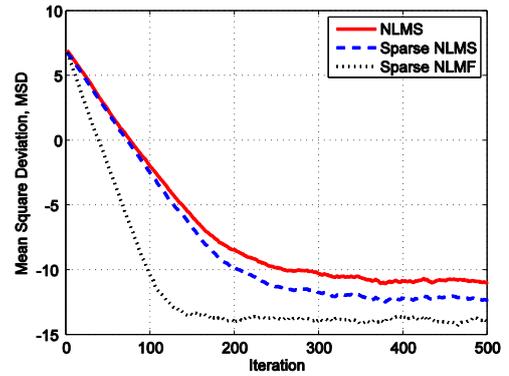

Fig. 5. MSD performances of NLMS, Sparse NLMS and Sparse NLMF algorithms.

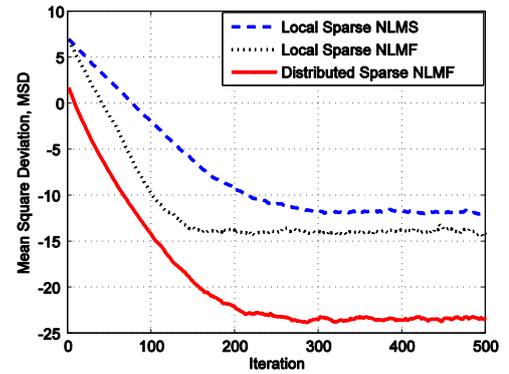

Fig. 6. MSD performances of Local Sparse NLMS, Local Sparse NLMF and Distributed Sparse NLMF algorithms.